\documentclass[amsmath,amssymb,aps,prc,twocolumn,superscriptaddress,showpacs,preprintnumbers]{revtex4}

\usepackage{graphicx,color}
\usepackage{multirow}
\usepackage{times}
\usepackage{bm}
\usepackage{epstopdf}
\usepackage{enumerate}

\usepackage[normalem]{ulem}  

\newcommand{\comment}[1]{}
\renewcommand\sout{\bgroup \color{red} \ULdepth=-.5ex \ULset}
\newcommand{\srtNN}{\sqrt{s_{{\scriptscriptstyle NN}}}}

\begin{document}
\preprint{YITP-17-80}

\title{
Enhancement of elliptic flow can signal a first order phase transition
in high energy heavy ion collisions
}

\author{Yasushi Nara}
\affiliation{
Akita International University, Yuwa, Akita-city 010-1292, Japan}
\affiliation{Frankfurt Institute for Advanced Studies, 
D-60438 Frankfurt am Main, Germany}
\author{Harri Niemi}
\affiliation{Institut f\"ur Theoretische Physik,
 Johann Wolfgang Goethe Universit\"at, D-60438 Frankfurt am Main, Germany}
\author{Akira Ohnishi}
\affiliation{Yukawa Institute for Theoretical Physics, Kyoto University,
Kyoto 606-8502, Japan}
\author{Jan Steinheimer}
\affiliation{Frankfurt Institute for Advanced Studies, 
D-60438 Frankfurt am Main, Germany}
\author{Xiaofeng Luo}
\affiliation{Key Laboratory of Quark\&Lepton Physics (MOE)
 and Institute of Particle Physics,\\
Central China Normal University, Wuhan 430079, China}
%
\author{Horst St\"ocker}
\affiliation{Frankfurt Institute for Advanced Studies, 
D-60438 Frankfurt am Main, Germany}
\affiliation{Institut f\"ur Theoretische Physik,
 Johann Wolfgang Goethe Universit\"at, D-60438 Frankfurt am Main, Germany}
\affiliation{GSI Helmholtzzentrum f\"ur Schwerionenforschung GmbH, D-64291
Darmstadt, Germany}

\date{\today}
\pacs{
25.75.-q, 
25.75.Ld, 
25.75.Nq, 
21.65.+f 
}

\begin{abstract} The beam energy dependence of the elliptic flow,
$v_2$, is studied in mid-central Au+Au collisions in the energy range of
$3\leq \srtNN \leq 30$ GeV within the microscopic transport model JAM.
The results of three different modes of JAM are compared; cascade-,
hadronic mean field-, and a new mode with modified equations of state,
with a first order phase transition and with a crossover transition.
The standard hadronic mean field suppresses the elliptic flow $v_2$, 
while the inclusion of the effects of
a first order phase transition (and also of a crossover transition)
does enhance the elliptic flow at $\srtNN<30$ GeV.
This is due to the high sensitivity of $v_2$ on the early,
compression stage, pressure gradients of the systems created
in high energy heavy-ion collisions.
The enhancement or suppression of the scaled energy flow, 
dubbed ``elliptic flow'', $v_2=\langle(p_x^2-p_y^2) / p_T^2\rangle$,
is understood as being due to
out of plane- flow, $p_y>p_x$, i.e. $v_2<0$,
dubbed out of plane - ``squeeze-out'',
which occurs predominantly in the early, compression stage.
Subsequently, the in-plane flow dominates,
$p_x > p_y$, in the expansion stage, $v_2 > 0$.
The directed flow, $v_1(y) = \langle p_x(y)/p_T(y)\rangle$,
dubbed ``bounce- off'',
is an independent measure of the pressure,
which quickly builds up the transverse momentum transfer in the reaction plane.
When the spectator matter leaves the participant fireball region,
where the highest compression occurs, 
a hard expansion leads to larger $v_2$.  
A combined analysis of the three transverse flow coefficients,
radial $v_0 \sim v_\perp$-, directed $v_1$- and elliptic $v_2$- flow,
in the beam energy range of
$3\leq\srtNN\leq10$ GeV,  distinguishes
the different compression and expansion scenarios:
a characteristic dependence on
the early stage equation of state is observed.
The enhancement of both the elliptic and the transverse
radial flow and the simultaneous collapse of the directed flow of nucleons
offers a clear signature if a first order phase transition is realized
at the highest baryon densities
created in high energy heavy-ion collisions.
\end{abstract}

\maketitle

\section{Introduction}

The determination of the QCD equation of state (EoS) is one of the most
important goals of high energy heavy-ion physics.
For this purpose,
the asymmetry of collective transverse flow 
in non-central nucleus-nucleus collisions,
reflected in the directed flow 
$v_1 = \langle\cos\phi\rangle$
and elliptic flow
$v_2=\langle\cos2\phi\rangle$
(where $\phi$ denotes the azimuthal angle of an outgoing particle
with respect to the reaction plane, and angular brackets denote
an average over particles and events), has been extensively studied.
In hydrodynamics, the final state momentum space asymmetry reflects the
pressure gradients in the early stage of the system
and is therefore sensitive to the EoS
~\cite{Stoecker:1980vf,Stoecker:1981pg,Buchwald:1984cp,
Stoecker:1986ci,Danielewicz:2002pu,Stoecker:2004qu}.
A (first-order) phase transition generally exhibits 
a softest point, i.e. a minimum in the speed of sound,
which is expected to have significant impact on
the flow observables~\cite{Hofmann:1976dy,Hung:1994eq}.
Indeed, a local minimum of the directed flow
~\cite{Rischke:1995pe,Brachmann:1999xt,Ivanov:2000dr}
with a negative rapidity-slope of directed nucleon flow
~\cite{Csernai:1999nf,Csernai:2004gk,Brachmann:1999xt,Brachmann:1997bq}
has been predicted by hydrodynamical simulations
in the case of first-order phase transition.
Recently, a negative rapidity slope of the proton directed flow
has been observed in the BES program
by the STAR Collaboration~\cite{STARv1,Shanmuganathan:2015qxb,Singha:2016mna}.
Although, a deeper understanding of the dynamics
of directed flow has been achieved
by recent more refined theoretical approaches
~\cite{Konchakovski:2014gda,Steinheimer:2014pfa,
Ivanov:2014ioa,Batyuk:2016qmb,Nara:2016phs,Nara:2016hbg},
a consistent theoretical interpretation of the 
experimental data on the excitation function of the directed flow
has not been obtained yet.

The elliptic flow $v_2$ has been measured by various experiments
from low to highest energies, currently available at LHC.
At low energies, from several hundred $A$MeV up to $\srtNN=4.3$ GeV,
a negative elliptic flow of protons, with respect to the reaction plane,
has been found
~\cite{Gutbrod:1988hh,E895v2,Andronic:2004cp,Andronic:2006ra,FOPI:2011aa}
which is called the 'squeeze-out'~\cite{Stoecker:1986ci}.
This predominant out-of-plane flow is due to the presence of
spectator matter, which blocks the participant particles escaping from
the collision zone~\cite{Danielewicz:2002pu,Danielewicz:1998vz}.
Therefore, a higher pressure due to a hard EoS leads to
a larger negative $v_2$~\cite{Hartnack:1994ce}.
See Ref.~\cite{LeFevre:2016vpp} for a recent
investigation on the detailed mechanism of the squeeze-out effect.
This squeeze-out can be reproduced by transport models which include
nuclear mean field, while cascade models cannot describe
this effect. This indicates that the kinetic pressure from ideal resonance gas
alone is too small. 
This lack of dynamical flow in the cascade type models
have been recognized for directed flow as well
~\cite{Molitoris:1986vm,Kahana:1994be}.
According to transport theoretical analysis 
in Ref.~\cite{Danielewicz:1998vz,E895v2,Danielewicz:2002pu},
the flow data up to top AGS energy $\srtNN<5$ GeV
can be described by repulsive mean field interactions at high density with 
the range of nuclear incompressibility of $K=220-380$ MeV.

On the other hand, at higher incident energies, where the passing
time of the spectators, $t_\text{pass}$, becomes
shorter, and the blocking effect of spectators becomes negligible,
the expansion of participant matter 
is stronger towards the in-plane direction.
Thus, the strength of $v_2$ is mainly determined by the
initial transverse overlap geometry (spatial eccentricity)
and the strength of the pressure gradient
~\cite{Ollitrault:1992bk}.
The passage time can be estimated by
$t_\text{pass}=2R/\gamma v$, where $R$ is the radius of the nucleus,
$\gamma$ and $v$ are the Lorentz factor
and the incident velocity of the colliding nuclei, respectively.
We expect the squeeze-out effect to become negligible above
$\srtNN\approx30$ GeV, since the passage time becomes less than 1 fm/c.
Thus, at highest energies, the higher pressure leads to
a stronger positive flow $v_2$, in contrast to
the beam-energy where the highest net-baryon densities are achieved
in heavy-ion collisions.
At RHIC and LHC, large elliptic flow values are found
~\cite{Ackermann:2000tr,Adcox:2002ms,Back:2002gz,Abelev:2014pua,Adam:2016izf},
which reach values compatible with ideal hydrodynamical predictions,
i.e.  with very small viscosities~\cite{Romatschke:2007mq,
Song:2010mg,
Gale:2012rq,
Niemi:2015qia,Niemi:2015voa}.
For recent reviews, see~\cite{Heinz:2013th,Gale:2013da,Huovinen:2013wma,
Hirano:2012kj,Jeon:2015dfa,Jaiswal:2016hex}.

This paper discusses
the collective flow in the beam energy region of the highest net-baryon
densities~\cite{Stoecker:1980uk}
between 5 and 20 GeV: Here an interesting situation emerges,
as pointed out in Ref.~\cite{Sorge:1996pc}:
The passage time of two Au nuclei is
$\sim$ 5 fm/c at $\srtNN=5$ GeV
and 1.25 fm/c at 20 GeV.
Thus, in this beam energy region, 
both the initial squeeze-out effect
(mostly during the compression stages of the reaction)
and the in-plane emission at the expansion stages
are important, and the interplay of them determines 
the final strength of the elliptic flow.
Ref.~\cite{Sorge:1996pc}
compared transport calculations with and without hadronic mean field
at $\srtNN=5$ GeV.
This work demonstrated that the final elliptic flow at mid-rapidity
is very sensitive to the pressure at maximum compression,
and it is conjectured that the
analysis of the elliptic flow can constrain the EoS in dense matter.

The beam energy dependence of elliptic flow has been
studied lately in distinctly different models
\cite{Auvinen:2013sba, Petersen:2006vm,
Petersen:2010md,
Karpenko:2015xea,
Steinheimer:2012bn,
Ivanov:2014zqa}:
The PHSD model predicts a smooth growth of the elliptic flow
with the beam energy as a result of increasing importance of
partonic degrees of freedom~\cite{Konchakovski:2012yg}.
This seems consistent with recent measurements
by the STAR collaboration~\cite{STARv2}.
The UrQMD hybrid model~\cite{Auvinen:2013sba}
finds that
the contribution of the hydrodynamical stage to $v_2$
is small at $\srtNN=5-7.7$ GeV.
Other UrQMD hybrid model approach, with viscous hydrodynamics,
reported that double the shear viscosity coefficient over entropy
density ratio, $\eta/s$,
is required to fit the elliptic flow at $7.7<\srtNN<11.5$ GeV
as compared to the higher beam energies $\srtNN>39$ GeV
~\cite{Karpenko:2015xea}.
A three-fluid dynamical (3FD) model simulation shows low sensitivity 
of $v_2$ for charged hadrons to the EoS~\cite{Ivanov:2014zqa}.

An important finding of the aforementioned model simulations is
the strong dependence of the final elliptic flow on the treatment of
the different stages of the collisions.
In most of the so-called hybrid models,
the different stages of the collisions are described by different approaches,
e.g, transport theory first, then hydro, then again transport theories
as after burner.
These particular prescriptions and the various matching
of the stages has substantial influence on the final observed elliptic flow.
Such a complicated treatment was deemed necessary
because most of the available standard hadronic transport models
seemed to be unable to describe the large values of the elliptic flow
experimentally observed at the top RHIC and LHC energies. 
The general paradigm has become that 
this insufficient description of the $v_2$ values
is due to a lack of interactions
in the dense phase, missing multi particle interactions,
as well as a lack of interactions between the pre-hadronic matter
consisting of unformed hadrons.

On the other hand, such shortcomings are considered to be less important
at beam energies which yield the highest net-baryon densities,
from  $\srtNN=3-30$ GeV. This opens the opportunity to study 
super dense quarkyonic and baryonic matter via measurements of
the elliptic flow in the beam energy range between 3 and 30 GeV,
as we will demonstrate with
the microscopic transport model JAM~\cite{JAMorg}:
This single, consistent approach,
used here does not require any sort of matching of different phases.
JAM has significantly reduced systematic uncertainties
studying the equation of state dependence of
the elliptic flow in the highest net-baryon densities.
The collective flows in mid-central Au+Au collisions are calculated
in three different scenarios within JAM:
a) the standard cascade mode~\cite{JAMorg},
b) the standard hadronic mean field mode~\cite{Isse},
and c) the cascade mode with a modified EoS~\cite{Nara:2016phs,Nara:2016hbg}.
These three modes of the JAM simulations
show different sensitivities of
the three transverse collective flow valuables,
the radial, the directed and the elliptic flow,
on the EoS, within a single consistent transport approach.

This paper is organized as follows:
Sec.~\ref{sec:model} describes the salient features of the JAM
transport approach.
Sec.~\ref{sec:result} presents the results for the excitation
function of the elliptic flow.
A detailed analysis of the collision dynamics is given in Sec.~\ref{sec:ana},
which demonstrates the importance of a combined analysis
of radial, directed, and elliptic flow.
The conclusions are given in Sec.~\ref{sec:conclusion}.

\section{The JAM Model}
\label{sec:model}

JAM~\cite{JAMorg} is a non-equilibrium microscopic transport model
based on the resonance and string degrees of freedom.
In JAM, particle production is modeled by the excitation and
decay of resonances and strings as employed by other models
~\cite{RQMD1995,UrQMD1,UrQMD2}.
Secondary products from decays of resonances or strings
can interact with each other
via binary collisions which generate collective flows.
A detailed description of the hadronic cross sections
and cascade method implemented in the JAM model can be found
in Ref.~\cite{JAMorg,Hirano:2012yy}.
In addition to the standard cascade simulation, the hadronic mean field
interaction has been incorporated in JAM~\cite{Isse,qm15no} based on the 
simplified version of relativistic molecular dynamics
(RQMD/S)~\cite{Maruyama:1996rn,Mancusi:2009zz}.
The RQMD approach~\cite{RQMD} is formulated to describe the covariant dynamics
of $N$ interacting particle system based on
the constrained Hamiltonian dynamics,
in which Hamiltonian is constructed from the sum of constraints, 
and the equations of motion are obtained by the condition that
the constraints must be conserved during the time evolution.
The RQMD/S formulation uses
the same mass shell conditions as the original RQMD formulation.
The difference is the choice of the simplified 
time fixation constraints which fix the time of the particles.
The time fixations are chosen so that
the time of all the particles are set to be the same in a reference frame.
If we further assume that the mean field is smaller than the kinetic
energy of the particles, which is justified in high energy collisions,
thus replacing the $p^0_i$ component as the kinetic energy in the
argument of the potentials,
one can solve for the Lagrange multipliers analytically.
As a result, the formulation is equivalent to the 
Hamiltonian system
\begin{equation}
  H = \sum_{i=1}^N \sqrt{\bm{p}_i^2 + m_i^2 + 2m_i V_i}
\end{equation}
with the equations of motion
\begin{align}
  \frac{d\bm{r}_i}{dt} &= \frac{\partial H}{\partial\bm{p}_i}
   =\frac{\bm{p}_i}{p^0_i}
   + \sum_{j=1}^N\frac{m_j}{p^0_j}
      \frac{\partial V_j}{\partial\bm{p}_j},\nonumber\\
  \frac{d\bm{p}_i}{dt} &= -\frac{\partial H}{\partial\bm{r}_i}
   = -\sum_{j=1}^N\frac{m_j}{p^0_j}
      \frac{\partial V_j}{\partial\bm{r}_j}.
\end{align}
Thus, the numerical cost of our approach is the same as
the non-relativistic  quantum molecular dynamics simulations.
RQMD/S approach has been successfully applied to the
various reactions~\cite{Isse,Mancusi:2009zz,Ogawa:2015zua}.
A similar approach is employed 
to describe the quark dynamics based on the NJL model~\cite{Marty:2012vs}.
We only consider a effect of hadronic mean field potentials in this work.
The effects of both partonic and hadronic mean fields
have been studied in the AMPT model~\cite{Xu:2016ihu}.

\begin{table}
\caption{Parameter set of the potentials in the mean field mode
 which yields the incompressibility of
$K=270, 370$ MeV in momentum-dependent soft (MS)
and hard (MH), respectively.  
$\rho_0=0.168$ 1/fm$^3$ is used for the normal
nuclear matter density.}
\begin{tabular}{ccccccccc}\hline\hline
type &$K$ & $\alpha$ & $\beta$ & $\gamma$ & $\mu_1$ & $\mu_2$  & $C_1$ & $C_2$
 \\
&(MeV) & (GeV)   & (GeV)   &   &  (1/fm) &   (1/fm) & (GeV) & (GeV) \\
 \hline
MS &270 & $-0.209$ & 0.284  & $7/6$    & 2.02  &   1.0   & $-0.383$ & 0.337 \\
MH &370 & $-0.0122$ & 0.0874 & $5/3$    & 2.02  &   1.0   & $-0.383$ & 0.337\\
\hline\hline
\end{tabular}
\label{table:potparam}
\end{table}

In the RQMD/S approach, particles acquire an effective mass 
$m^*_i = \sqrt{m_i^2 + 2m_iV_i}$
as a result of the interaction with the other particles,
where the scalar one-particle potential $V_i$ is given by
the sum of the Skyrme-type, density dependent terms and the Lorentzian-type,
momentum dependent terms:
\begin{align}
V_i &= \frac{\alpha}{2\rho_0}\langle\rho_i\rangle
    +\frac{\beta}{(1+\gamma)\rho_0^\gamma}\langle\rho_i\rangle^\gamma
      \nonumber\\
  &+\sum_{k=1,2} \frac{C_k}{2\rho_0}\sum_{j(\neq i)}
         \frac{1}{1+[p_{ij}/\mu_k]^2}\rho_{ij}\,.
\label{eq:pot}
\end{align}
Here,
\begin{equation}
\langle\rho_{i}\rangle=\sum_{j(\neq i)}\rho_{ij}
         =\sum_{j(\neq i)}(4\pi L)^{-3/2}\exp(q^2_{ij}/4L)
\end{equation}
and $q^2_{ij}$ and $p^2_{ij}$
are the relative distance and momentum squared
between particles $i$ and $j$ in their center of mass frame.
The Gaussian width is taken to be $L=1.08$ fm$^2$~\cite{BQMD}.
Potential parameters are determined by fulfilling the 
nuclear saturation properties and
the global optical potential~\cite{Hama:1990vr}.
In this paper, the parameter set listed in Table~\ref{table:potparam}
is employed.
The parameter set MS is used throughout this paper except for
the results in Table~\ref{table:v012}.

\begin{figure}[tbh]
\includegraphics[width=8.7cm]{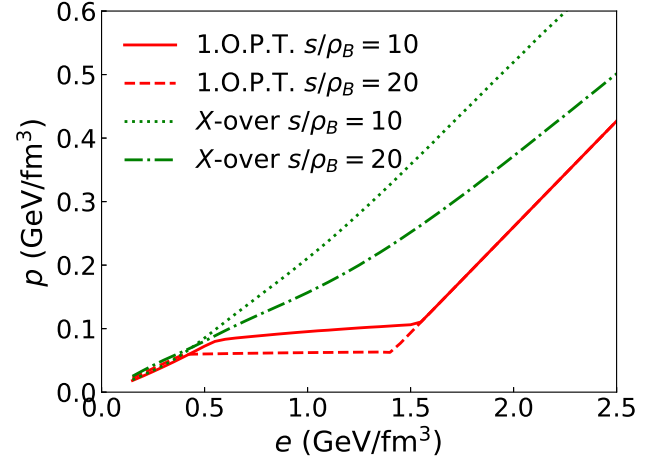}
\caption{EoS implemented in JAM.
The sold and dashed lines represent the EoS with a first order 
phase transition for the ratio of entropy to baryon density
$s/\rho_B=10$ and 20 respectively,
and the dotted and dotted-dash lines
represents the EoS with a crossover for $s/\rho_B=10, 20$.
}
\label{fig:eosstatic}
\end{figure}

\begin{figure*}[t]
\includegraphics[width=8.0cm]{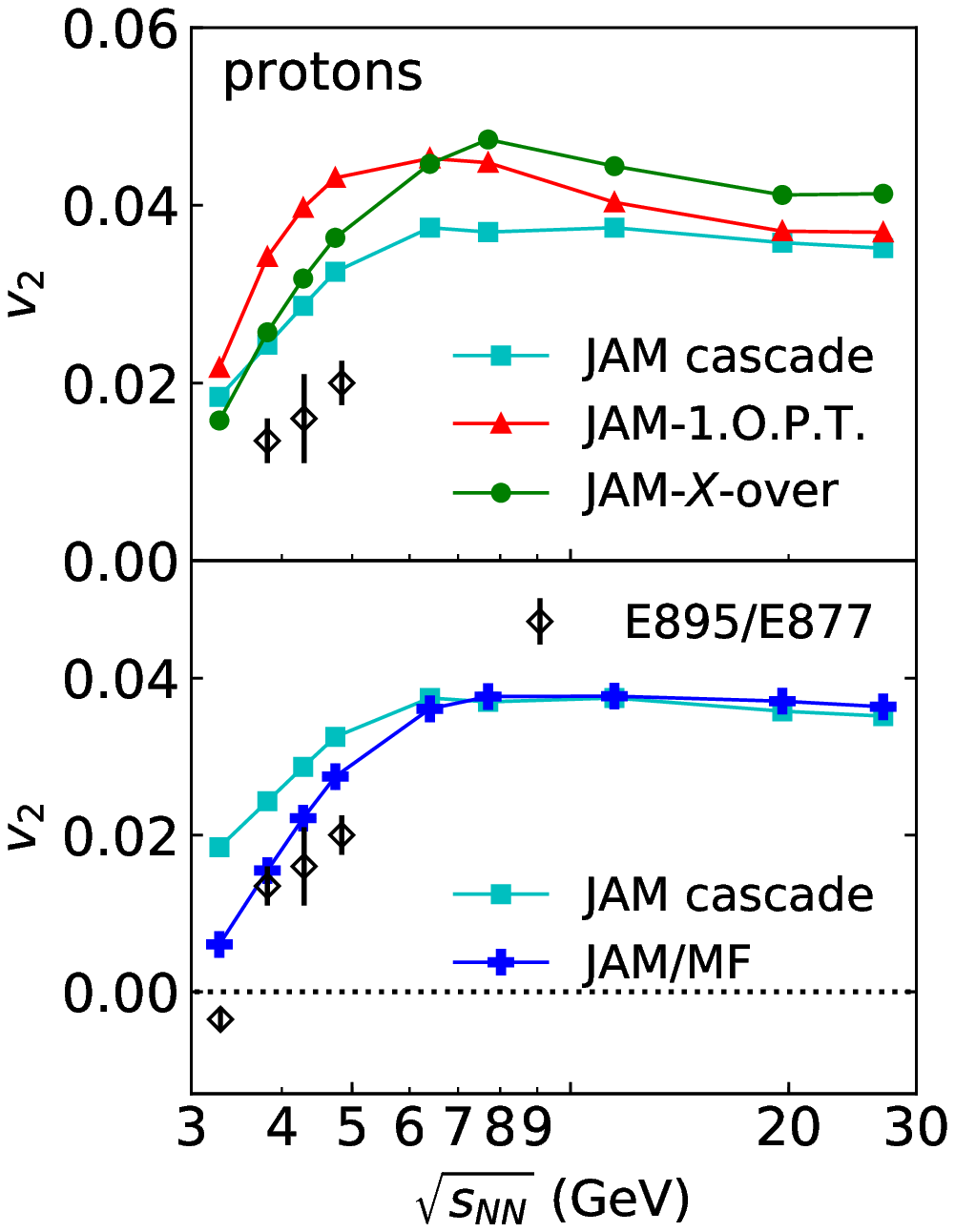}
\includegraphics[width=8.0cm]{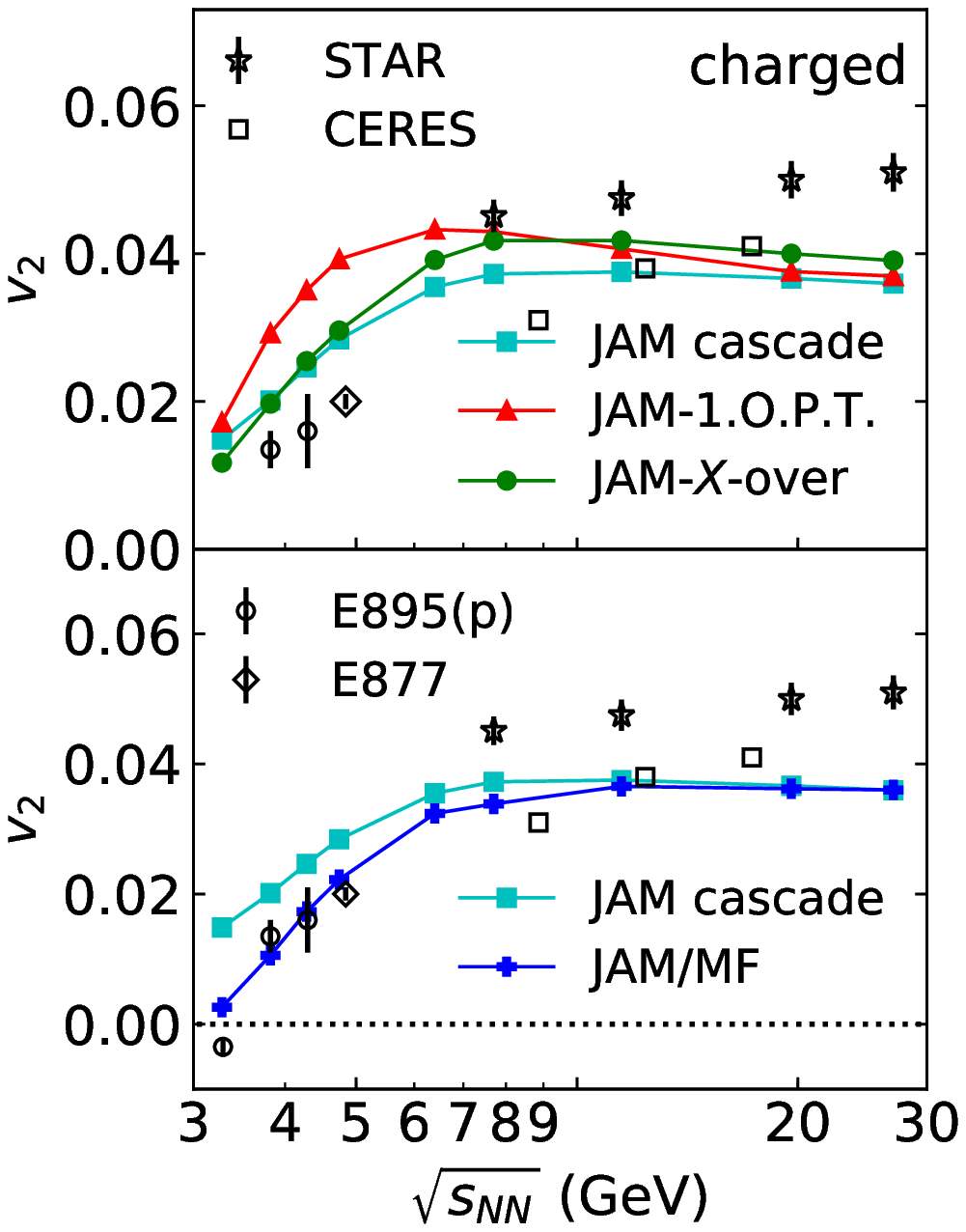}
\caption{Beam energy dependence of the elliptic flows
$v_2$ of protons at $|y|<0.2$ (left)
and charged hadrons at $|\eta|<0.2$(right)
in mid-central Au+Au collisions ($4.6\leq b\leq9.4$ fm)
from JAM cascade mode (full squares),
JAM with first-order EoS (full triangles),
crossover EoS (full circles),
and JAM with hadronic mean field (full crosses).
Data are from the E895/E877~\cite{Filimonov:2001fd,E895v2},
CERES~\cite{CERES2002},
and STAR~\cite{STARv2}
collaborations.
}
\label{fig:v2exfunc}
\end{figure*}

We also study the effect of the EoS on the elliptic flow
by employing the method proposed by Ref.~\cite{Sorge:1998mk}
based on the virial theorem~\cite{Danielewicz:1995ay},
in which the scattering style
of the two-body collisions are modified so as to control
the pressure of the system.
In this method, 
the azimuthal angle of the two-body scattering between particle $i$ and $j$
is constrained by~\cite{Nara:2016hbg}
\begin{equation}
 \Delta P = \frac{\rho}{3(\delta\tau_i+\delta\tau_j)}
                 \Delta\bm{p}_{ij}\cdot\Delta\bm{r}_{ij},
\label{eq:pre}
\end{equation}
where $\rho$ is the Lorentz invariant local particle density
and $\delta\tau_{i}$ is the proper time interval between
successive collisions.
$\Delta\bm{p}_{ij}$ is the momentum transfer
and $\Delta\bm{r}_{ij}=\bm{r}_i-\bm{r}_j$
is the relative coordinate between particle $i$ and $j$
in the two-body c.m. frame.
$\Delta P$ is the deviation of the pressure from the ideal gas value.
An advantage of this method is to provide a numerically efficient way
to modify the EoS of a system according to a given EoS table.
We implemented the effects of
both the EoS with a first-order phase transition (JAM-1.O.P.T.)
and a crossover transition (JAM-$X$-over)
based on this method~\cite{Nara:2016hbg}.
For the first-order phase transition EoS, 
we use the same model as the so-called EOS-Q~\cite{EOSQ}.
For the crossover ($X$-over) EoS, 
we use the chiral model EoS from Ref.~\cite{CMEOS}.
Here the EoS at vanishing baryon density is consistent
with a smooth crossover transition, as found in lattice QCD.
That EoS predicts two critical points at 
$\mu_B\approx 900$ and 1400 MeV.
The EoS from both models,
EoS-Q and $X$-over implemented in this work are compared
in Fig.~\ref{fig:eosstatic}
for entropy to baryon density $s/\rho_B=10$ and 20.

\section{Results}
\label{sec:result}

In the following we will present computations of the elliptic flow
in mid-central Au+Au collisions, and
compare the hadronic mean field results with the two types of
EoS described above.
In the simulation, we choose the mid-impact parameter range
$4.6<b<9.4$ fm, which is reasonable for mid-central collisions.

\subsection{Elliptic flow excitation functions}

Fig.~\ref{fig:v2exfunc} shows
the beam energy dependence of the elliptic flow $v_2$
of protons at mid-rapidity $|y|<0.2$, (left panel),
and charged particles, ($|\eta|<0.2$) (right panel),
in mid-central Au+Au collisions using the standard JAM cascade mode,
JAM with a first-order phase transition (JAM-1.O.P.T.),
JAM with a crossover EoS (JAM-$X$-over)
and JAM with hadronic mean field (JAM/MF),
compared to data from the STAR, CERES, and E895/E877 collaborations
~\cite{STARv2,CERES2002,Filimonov:2001fd,E895v2}.
JAM in the cascade mode does not reproduce the $v_2$-values
in the measured beam energy region:
The calculations overpredicts $v_2$ below $\srtNN=5$ GeV,
while it underpredicts $v_2$ above $\srtNN=7.7$.
At lower AGS energies, $\srtNN<5$ GeV,
the inclusion of nuclear mean fields reduces the $v_2$ values significantly,
and the JAM/MF mode calculations are in reasonable agreement
with the data, consistent with previous work
~\cite{Danielewicz:1998vz,E895v2,Danielewicz:2002pu,UrQMD1}.
Thus, high pressure in the early stages of heavy-ion collision achieves
a stronger squeeze-out at low beam energies.
Certain model uncertainties prevail
related to the implementation of mean field:
In the RQMD/S approach, potentials are included as Lorentz scalars.
However, vector potential yield stronger repulsion effects than
scalar potentials.
In fact, the present model yields slightly larger $v_2$ values than
seen in predictions by the BUU-
~\cite{Danielewicz:1998vz,E895v2,Danielewicz:2002pu}
and UrQMD model~\cite{UrQMD1},
in which vector potentials are implemented.
Influence of vector potentials on $v_2$ is checked
by simulating
the potential in Eq.(\ref{eq:pot})
as the zeroth component of a vector potential~\cite{NiitaQMD}.
It is found that $v_2$ is more suppressed, up to AGS energies
(about a 20\% reduction at $\srtNN=4.74$ GeV),
but that effect is small at higher energies.

In contrast, calculations with a soft EoS,
such as JAM-1.O.P.T. and JAM-$X$-over, yield larger
$v_2$ values, which lead to smaller early pressure gradients,
resulting in less squeeze-out.
All calculations lead to similar $v_2$ values
at high energies.
The calculated $v_2$ values here are lower than the data.
This may indicate the need to include parton dynamics,
as conjectured by PHSD simulations~\cite{Konchakovski:2012yg},
where the dynamics of the partonic phase becomes relevant
at beam energies above $\srtNN=10$ GeV.
Hybrid models predict comparable
$v_2$ values, above $\srtNN=10$ GeV~\cite{Karpenko:2015xea},
if a hydrodynamic phase with small viscosity is included in the simulation.

Our results suggest that the intermediate energy region,
$5<\srtNN<7.7$ GeV, is most interesting:
If there is no softening of the EoS,
and the hadronic pressure dominates, $v_2$ must be suppressed
to nearly the same values as in the cascade calculations.
On the other hand, if the softening of the EoS
plays an important role, then $v_2$-values are enhanced.

Strong initial non-equilibrium pressure due to the hadronic mean-field
explains the suppression of the elliptic flow
at AGS energies. 
The dynamical treatment of a first order chiral phase transition in a
non-equilibrium real-time dynamics
predicts strong reduction of elliptic flow~\cite{Paech:2003fe}.

\subsection{Event-plane method and Cumulants}

Various methods have been proposed to extract the anisotropic flows.
As the reaction plane is not a priori known in heavy ion experiments,
different methods on how to extract elliptic flow must be tested.
First, elliptic flow $v_2\{\mathrm{EP}\}$ is computed
by the event-plane method~\cite{Poskanzer:1998yz},
where the anisotropic flow coefficients $v_n\{\mathrm{EP}\}$ are given by
\begin{equation}
 v_n\{\mathrm{EP}\} = \frac{\langle\cos[n(\phi - \Psi_n)]\rangle}
                  {\sqrt{2\langle\cos[n(\Psi_n^A-\Psi_n^B)]\rangle}}.
\label{eq:vnep}
\end{equation}
The event plane angle $\Psi_n$ is then estimated from the event flow vector
$Q_n$ via
\begin{equation}
 \Psi_n = \arg (Q_n)/n,~~~~~
 Q_n=\sum_j w_j e^{in\phi_j},
\end{equation}
where the sum runs over all particles used in the event-plane determination
except for the particle actually used for the evaluation of $v_2$,
to remove self-correlations.
Here particles with $1<|\eta|<3$ are used to avoid self-correlations
in the calculations of $v_2$ at mid-rapidity.
$\phi_j$ is the azimuthal angle
for the $j$-th particle in the momentum space.
The transverse mass of particle $i$ is used as the weight $w_j=m_{Tj}$.
Two subevents, $A$ and $B$, are used to estimate the event plane
resolution in the denominator of Eq.~(\ref{eq:vnep}). 
The two subevents are defined - in the forward ($A$: $-3<\eta<-1$)
and - in the  backward ($B$: $1<\eta<3$) regions.

Another standard method is the cumulant method:
here multi-particle correlations are used to extract
the anisotropic flow parameters~\cite{Borghini:2000sa,Bilandzic:2010jr}.
The scalar product method~\cite{Adler:2002pu} is used in order to suppress
the so-called non-flow effects, by imposing pseudo-rapidity gap $|\Delta\eta|$.
This method separates the particles in the mid-rapidity region
into two subevents, $a$ and $b$, each with multiplicity,
$M_a$ and $M_b$, which are separated by a rapidity gap.
The formula\cite{Adler:2002pu,Zhou:2015eya}
\begin{equation}
v_n\{2,|\Delta\eta|\} = \sqrt{\frac{\langle Q_n^a Q_n^{b*}\rangle}
                    {\langle M_a M_b \rangle}}
\end{equation}
now computes $Q_n^{a/b}$ from  the particles
in the two regions respectively, as separated by the gap $|\Delta\eta|$
in the event with the weight $w_j=1$.

\begin{figure}[tbh]
\includegraphics[width=8.5cm]{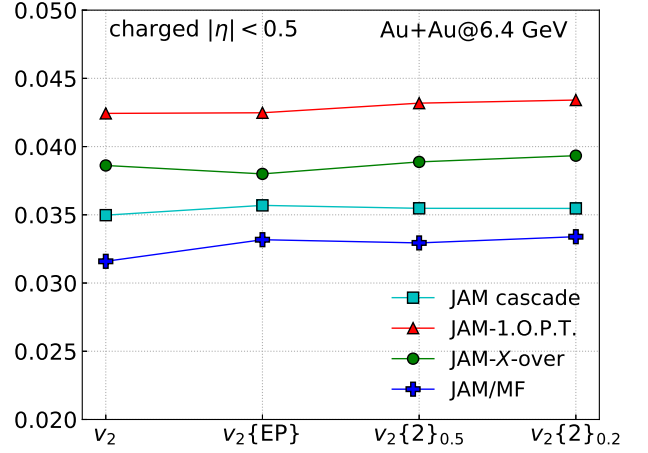}
\caption{Comparison of elliptic flow at mid-rapidity ($|\eta|<0.5$)
extracted by different methods
in mid-central Au+Au collisions at $\srtNN=6.4$ GeV
for cascade mode (squares),
first-order EoS (triangles),
crossover EoS (circles),
and hadronic mean field mode (crosses).
$v_2\{2\}_{0.5}$ and $v_2\{2\}_{0.2}$ 
correspond to the results with pseudo-rapidity gaps of
$|\Delta\eta|>0.5$ and $|\Delta\eta|>0.2$ respectively.
}
\label{fig:v2ch64ec}
\end{figure}

Fig.~\ref{fig:v2ch64ec} compares elliptic flows
of charged hadrons at mid-rapidity, $|\eta|<0.5$,
extracted by these different methods
in mid-central Au+Au collisions at $\srtNN=6.4$ GeV.
Fig.~\ref{fig:v2ch64ec} shows
different values for the rapidity gap denoted by
$v_2\{2\}_{0.5}$ and $v_2\{2\}_{0.2}$,
corresponding to $|\Delta\eta|>0.5$ and $|\Delta\eta|>0.2$.
Observe that both, event-plane and cumulant methods, are close
to the reaction plane elliptic flow $v_2$ value for all modes in JAM.
The results are similar for other beam energies.
Hence, these conclusions are robust with respect to the different methods
used to extract the elliptic flow.
The STAR collaboration~\cite{STARv2} reported
that the difference between four-particle cumulants $v_2\{4\}$
and $v_2\{2\}$ reduces at the lower collision energies,
and that $v_2\{4\}\approx v_2\{2\}\approx v_2\{\mathrm{EP}\}$,
for mid-central collisions at $\srtNN\leq 11.5$.
Our results are consistent with these observations.

The next section examines in detail the collision dynamics,
in particular how the elliptic flow is generated during the reaction.

\subsection{Analysis of the collision dynamics}
\label{sec:ana}

\begin{figure}[tbh]
\includegraphics[width=7.5cm]{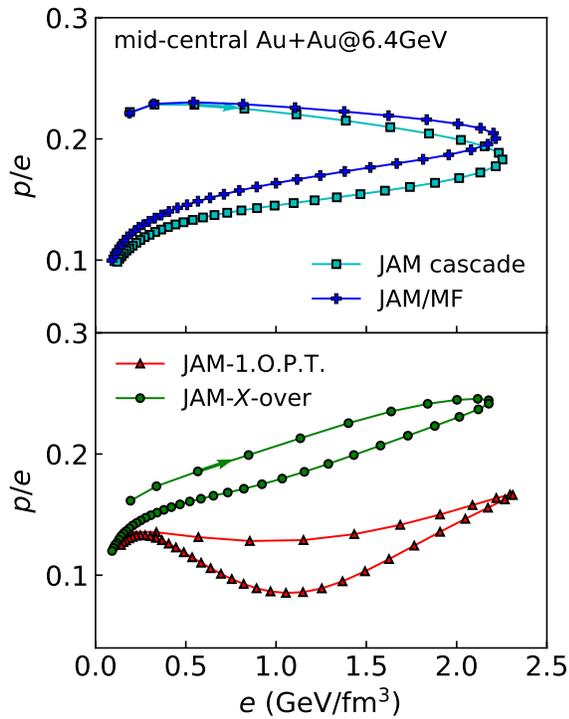}
\caption{Time evolution of the isotropic pressure $p$,
and energy density $e$
in mid-central Au+Au collisions at $\srtNN=6.4$ GeV
calculated in the JAM cascade mode (squares),
first-order EoS (triangles),
crossover EoS (circles),
and hadronic mean field mode (crosses).
Initial points correspond to the time 0.375 fm/c, after the touching
of the two nuclei, and the time interval is 0.25 fm/c.
Arrows indicate the direction of the time evolution of the reactions.
Pressure and energy density
are averaged over collision points in a cylindrical volume of transverse
radius 3 fm and a longitudinal distance of 2 fm centered at the origin.
}
\label{fig:eos64}
\end{figure}

The sensitivity of the elliptic flow
on the early pressure gradients,
i.e. the EoS is analysed 
in Fig.~\ref{fig:eos64}, 
by examining the time evolution of
the local isotropic pressure $p$,
and the energy density $e$, averaged over events,
in mid-central Au+Au collision at $\srtNN=6.4$ GeV.
The results are compared for the different EoS employed  in JAM.
These quantities are averaged values over collision points,
thus those particles which have not yet collided are not included in the 
evaluation.
We evaluate the non-ideal pressure by using Eq.~(\ref{eq:pre}),
as in Ref.~\cite{Nara:2016hbg}, for both the JAM-1.O.P.T. 
and the $X$-over mode.
For the potential interactions in the JAM/MF mode,
the pressure from the potential contribution
$\Delta P_\text{pot}$
is given locally by~\cite{Sorge:1996pc}
\begin{equation}
 \Delta P_\text{pot}=\frac{\rho}{3}(\bm{F}_i\cdot\bm{r}_i
           +\bm{F}_{ri}\cdot\bm{p}_i)\,.
\end{equation}
Here $\rho$ is the particle density, 
$\bm{F}_i$ and $\bm{F}_{ri}$ are the forces
exerted on the $i$-th particle due to the density-
and the momentum-dependent parts of the potentials, respectively.
The JAM/MF-EoS is stiffer than the EoS in the cascade mode,
due to the density- as well as momentum-dependent potentials
which predict repulsive interactions at high baryon densities.
The mean field effects are
more pronounced at lower energies in JAM/MF.
In stark contrast to the JAM/MF mode,
the EoS with the first-order phase transition (JAM-1.O.P.T.)
predicts very low $p/e$ values, indicating a small sound velocity;
while the crossover EoS is softer than that of
the cascade mode, in the early stages of the collisions.
Just this EoS becomes the stiffest EoS at the maximum overlap of the two nuclei.
The origin of the hardness of the EoS is similar
between JAM/MF and JAM-$X$-over:
In JAM/MF, the EoS is hard because of the repulsive hadronic interactions,
in the crossover EoS the dense hadronic part of the EoS is stiff
due to hard core repulsion effects.
Bag model massless ideal gases of quarks and gluons
is used in the JAM-1.O.P.T.. Hence, the QGP phase has the same
$p$ and $e$ dependence, independent of the baryon density.
After the system expands and reaches the hadronic phase,
all EoS are harder than the EoS in the JAM cascade mode,
due to the repulsive potentials implemented in the EoS.

\begin{figure}[tbh]
\includegraphics[width=7.5cm]{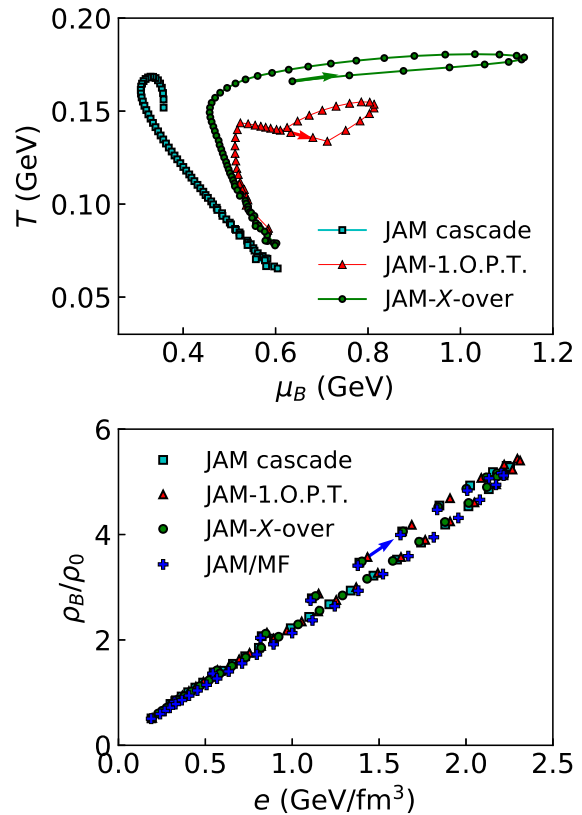}
\caption{Time evolution of temperature $T$ and baryon chemical potential
$\mu_B$ (upper panel), and
net-baryon density $\rho_B/\rho_0$,
and energy density $e$ (lower panel)
in mid-central Au+Au collisions at $\srtNN=6.4$ GeV.
}
\label{fig:eos64tmu}
\end{figure}

Fig.~\ref{fig:eos64tmu} displays the time evolution of
the temperature $T$, the baryon chemical potential $\mu_B$,
the net-baryon density $\rho_B/\rho_0$, and the energy density $e$
in mid-central Au+Au collision at $\srtNN=6.4$ GeV.
All modes yield similar behavior for both, the net-baryon density
and the energy density. However, the $T$-$\mu_B$ evolutions are quite
different among different EoS.
Both, the first order phase transition and the crossover transition,
are clearly seen by the trajectories in the $T$-$\mu_B$ plane.
The crossover EoS in JAM follows similar trajectory as
an isotropic expansion using
the lattice QCD calculations~\cite{Bellwied:2016cpq}.

Besides the uncertainty in the EoS, 
there is another model degree of freedom in the JAM EoS-modified mode,
where the pressure is controlled by imposing
a static equilibrium EoS $p(e,\rho_B)$.
This prescription may be inadequate
when the system is in a highly non-equilibrated state,
as in particular, during the very early stages of the collisions.
Keep this feature in mind for the analysis below.
This problem does not exist in the mean field
approach, which treats the effect of the EoS naturally
in the non-equilibrium stages of the collisions.

\begin{figure}[t]
\includegraphics[width=8.0cm]{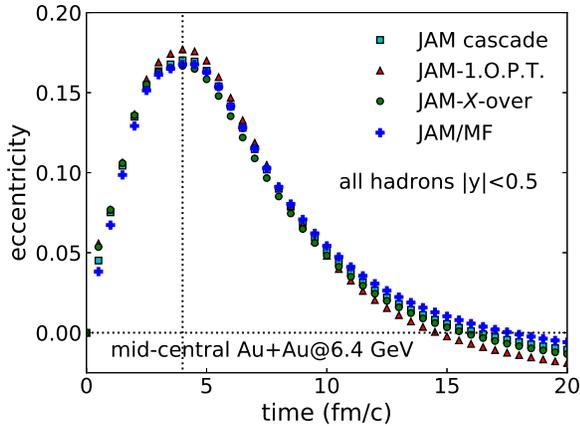}
\caption{
The time evolution of the eccentricity $\epsilon$ as evaluated from
all particles in mid-rapidity $|y|<0.5$.
}
\label{fig:ecc}
\end{figure}

The elliptic flow is sensitive to the initial deformation of
the almond shape participant, geometry representing
in non-central collisions:
Fig.~\ref{fig:ecc} shows the time evolution of the eccentricity
$\epsilon$ of the system in Au+Au collision at $\srtNN=6.4$ GeV
defined as
\begin{equation}
\epsilon = \left\langle\frac{y^2-x^2}{x^2+y^2}\right\rangle.
\end{equation}
Note that the eccentricity of the system reaches a maximum
at time $t=4$ fm/c, the passage time
of the spectators for this incident energy.
Thus, the role of the pressure flips
at this time, $t=4$ fm/c: higher pressure leads to
a reduction of $v_2$ at $t<4$ fm/c,
while higher pressure leads to an enhancement of $v_2$ at $t>4$ fm/c.
Observe that JAM-1.O.P.T. yields 
the largest eccentricity caused by the compression of the system.
This large eccentricity generates large $v_2$-values
even though the JAM-1.O.P.T. mode
gives the lowest pressure among the different modes.

\begin{figure}[t]
\includegraphics[width=8.0cm]{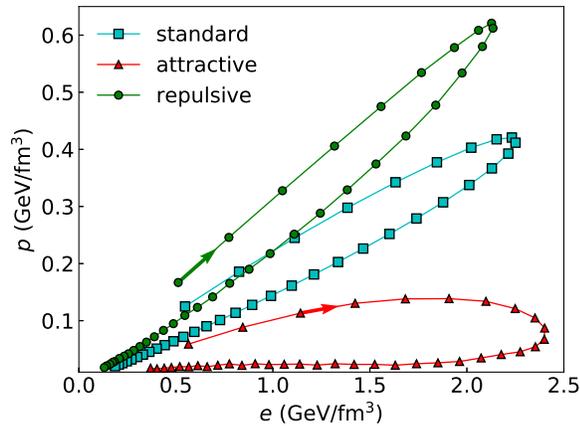}
\caption{Time evolution of the local pressure $p$
and energy density $e$ in mid-central Au+Au collision 
at $\srtNN=6.4$ GeV for JAM standard cascade simulation (squares),
JAM with attractive orbit (triangles),
and JAM with repulsive orbit (circles).
}
\label{fig:attrep}
\end{figure}

The interplay between the initial pressure
and late pressure, (initial and late pressure is defined as the
pressure before or after the passage time $t_\text{pass}$ of the spectators),
is studied by performing
simulations which impose either attractive or repulsive
orbits for all two-body collisions.
Changing this ``collision style'' is known to change the pressure of the system.
Selecting attractive orbits in a two-body scattering reduces the pressure.
On the other hand, repulsive orbits enhance the pressure.

\begin{figure}[t]
\includegraphics[width=8.0cm]{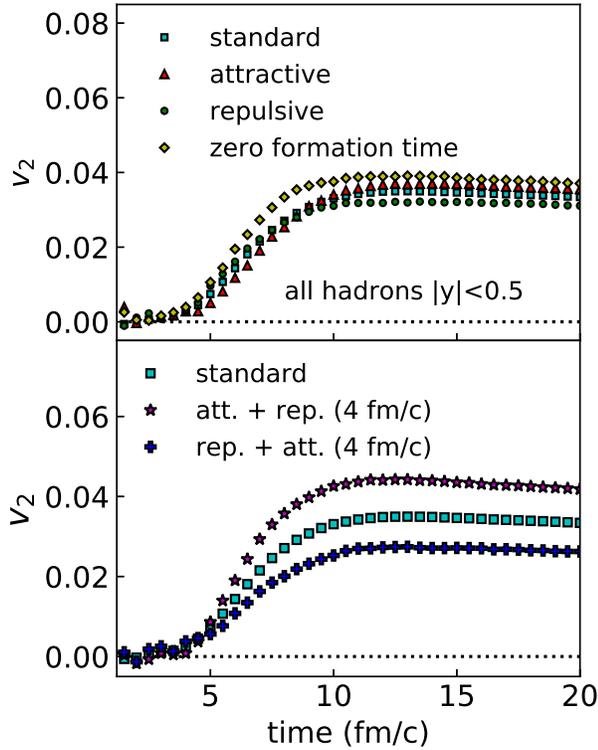}
\caption{
Upper panel:The time evolution of $v_2$ in
mid-central Au+Au collision at $\srtNN=6.4$ GeV 
for JAM standard cascade simulation (squares),
JAM with attractive orbit (triangles),
and JAM with repulsive orbit (circles).
Diamonds show the results from the JAM standard cascade mode
with zero formation time.
Lower panel:
JAM with attractive (repulsive) orbit
before the passing of spectators time
$t<4$ fm/c and repulsive (attractive) orbit at $t>4$ fm/c
are shown by stars (crosses).
}
\label{fig:v2pres}
\end{figure}

The time evolution of the effective EoS for each simulation
is shown in Fig.~\ref{fig:attrep}.
Note that the pressure is modified by an order of magnitude
in these simulations.
As expected, simulations with attractive orbits significantly
soften the EoS, while, repulsive orbit simulations
show extremely high pressure values.
The time evolution of $v_2$ for these particular cases
are shown in Fig.~\ref{fig:v2pres}.
The reduction of $v_2$, due to initial high pressure
in the repulsive orbit simulation,
is partially canceled by the enhancement of the $v_2$
during the late expansion stages.
Note that the same cancellation occurs in the attractive orbit simulation.
In our previous work~\cite{Nara:2016phs},
attractive orbit simulations yielded the 
nearly the same $v_2$-values as the JAM cascade mode
for charged hadrons from $\srtNN=7.7$ GeV up to  $\srtNN=27$ GeV.
This is understood as being due to the cancellations on $v_2$.
On the other hand, attractive orbit simulations
at lower incident energies, $\srtNN \leq 7$ GeV,
yield an enhancement of $v_2$,
because the initial pressure dominates the final value of elliptic flow
at such low beam energies~\cite{Chen:2017cjw}.

The role of initial and final pressures is demonstrated
in Fig.~\ref{fig:v2pres} by increasing the rate of two-body hadron-hadron
collisions in the simulation:
Here, the cascade simulation uses  zero hadron formation time,
to simulate faster equilibration with highest pressure
by increasing the number of collision.
Still, one notes the partial cancellation even in these calculations.
Zero formation time simulations using UrQMD at $\srtNN=200$ GeV
reach the ideal hydro limit~\cite{Bleicher:2000sx}.
This cancellation may be the reason why 3FD model simulations
show an insensitivity of $v_2$ to the EoS~\cite{Ivanov:2014zqa}
at the STAR BES energies.

The compensating effects of early and late pressures are explicitly
confirmed by a simulation in which attractive orbits are selected
\textit{before} the passage time $t_\text{pass}=4$ fm/c, 
but repulsive orbits are selected
\textit{after} the passing of the spectator nucleons:
Here, a strong enhancement of $v_2$ is expected, and, vice versa,
$v_2$ should drop, if repulsive orbits are employed
\textit{before} $t_\text{pass}$,
and attractive orbits \textit{after} $t_\text{pass}$.
Indeed, the predicted enhancement of $v_2$ as well as the reduction of $v_2$
are manifested in such simulations, 
see Fig.~\ref{fig:v2pres} (stars and crosses).
This proves that
the final values of the elliptic flow
are due to the (partial) cancellation of the early
squeeze-out with the subsequent in-plane bounce off flow
~\cite{Stoecker:1981pg}.
This analysis allows to summarize the different stages of
the collision dynamics leading to the different values of $v_2$
at beam energies below $\srtNN \approx 10$ GeV as follows:

\begin{enumerate}[(i)]
\item For the hadronic mean field case,
elliptic flow is suppressed-because the initial squeeze-out due to
repulsive potentials is stronger than the late in-plane flow,
for beam energies $\srtNN \lesssim 5.5$ GeV.
At higher beam energies,
elliptic flow is nearly the same in the MF case as in the cascade mode,
due to cancellations and less effect of the potentials.

\item For the first-order phase transition,
the system is mostly inside of the soft region in this beam energy range, 
and the pressure is always very low.  This leads to
high compression and to a very soft expansion of the system.
In addition, this softening generates large eccentricities.
As a consequence, a first-order phase transition
largely suppresses the initial squeeze-out. 
Hence, it yields large positive $v_2$-
even though the pressure is low.

\item For the crossover transition,
the pressure is soft in the initial compression stage, but
it becomes high in the expansion stage.
As a result, the final elliptic flow is large.
\end{enumerate}

\subsection{Radial, directed and elliptic flows}

\begin{figure}[t]
\includegraphics[width=8.0cm]{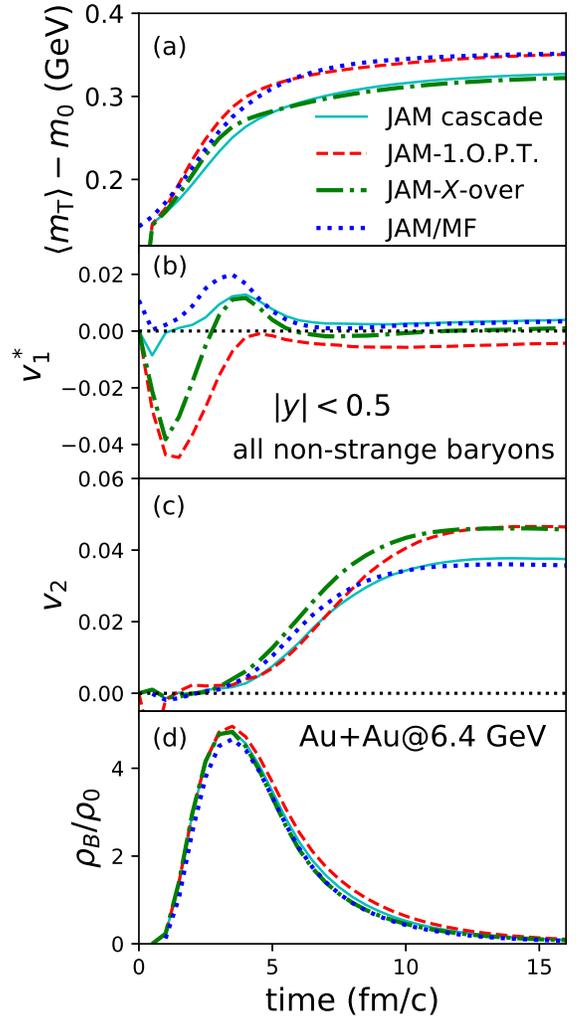}
\caption{The time evolution of
(a)  average transverse mass
$\langle m_\textrm{T}\rangle-m_0$,
(b) sign weighted directed flow $v_1^*$,
(c) elliptic flow $v_2$ for baryons
at mid-rapidity $|y|<0.5$
in mid-central Au+Au collision at $\srtNN=6.4$ GeV.
Time evolution of normalized baryon density $\rho_B/\rho_0$ 
averaged over a cylindrical volume of transverse radius 3 fm
and longitudinal distance of 1 fm 
is displayed in the panel (d).
Solid, dashed, dashed-dotted, and dotted lines show the results from
JAM cascade, JAM with first-order phase transition,
JAM with crossover EoS,
and JAM with hadronic mean field simulations,
respectively.
}
\label{fig:v2evol}
\end{figure}

Turning to the combined analysis of the three different flow coefficients,
namely the radial, directed and elliptic flow,
distinction of different effects of the EoS on the different flow coefficients
are observed.
This independent sensitivities allow for distinguishing
the different expansion dynamics.

Fig.~\ref{fig:v2evol} displays the time evolution of the
average transverse mass
$\langle m_\mathrm{T}\rangle-m_0$, where
$m_\mathrm{T}=\sqrt{\bm{p}^2_\mathrm{T}+m_0^2}$,
the sign weighted directed flow,
\begin{equation}
v_1^* = \int dy\, v_1(y)\textrm{sgn}(y),
\end{equation}
and the elliptic flow $v_2$,  of baryons
at mid-rapidity, $|y|<0.5$,
as well as the normalized net-baryon density,
in Au+Au collision at $\srtNN=6.4$ GeV
for four different cases: Here,
$\langle m_\textrm{T}\rangle -m_0$
characterizes the strength of the radial transverse flow.
Both JAM-1.O.P.T. and JAM/MF simulations yield
the enhanced $\langle m_\textrm{T}\rangle$.
However, the dynamical origin of the large radial flow is different:
The radial flow in JAM/MF is generated
due to the hadronic repulsive potentials,
while in JAM-1.O.P.T. it is the result of 
the stronger compression of the system, and
the longer lifetime of the system.
We note that the difference of transverse dynamics between
the first order phase transition and the cross over transition
is consistent 
in our simulations
with the predictions by 
the hydro+UrQMD approach~\cite{Petersen:2009mz}.

For the directed flow $v_1$, the softening of the EoS leads to 
a dramatic effect: negative $v_1$ values as experimentally observed,
for baryons in JAM-1.O.P.T.
In stark contrast, the cascade, the crossover, and the mean field
simulations predict a positive $v_1$-slope.
The $v_1$ value and the sign are very sensitive to the EoS.
It is important to study in detail the EoS dependence on the net-baryon
density by $v_1$ in the future.
The elliptic flow $v_2$ develops in time scales compatible with
the system size/$c$; hence, $v_2$ increases between 5 fm/c and 10 fm/c.
The excitation function of the elliptic flow
shown in Fig.~\ref{fig:v2exfunc} demonstrated that
JAM/MF yields nearly the same amount of $v_2$
as the cascade mode, while
JAM-1.O.P.T. and JAM-$X$-over predict larger $v_2$ values.

\begin{table}[t]
\caption{Slope parameter $T_\textrm{eff}$, 
the slope of directed flow $dv_1/dy$ and the elliptic flow $v_2$
of mid-rapidity ($|y|<0.2$)
nucleons in mid-central Au+Au collision at $\srtNN=6.4$ GeV.
The comments in the parenthesis are relative to the cascade mode.
Scalar type momentum dependent hard (MH/Scalar) and soft (MS/Scalar) mean field,
and vector type momentum dependent hard (MH/Vector) and soft (MS/Vector)
mean field
are also compared.
}
\begin{tabular}{lrrr}\hline\hline
         & $T_\textrm{eff}$ (MeV)
  & $v_1$ slope (\%)      & $v_2$  (\%)\\\hline
cascade     &  259             & $2.46$             & 3.74  \\
1st-order   &  278 (enhanced)  & $-0.34$ (negative) & 4.45 (enhanced)  \\ 
crossover   &  257 (same)      & $1.52$ (positive) & 4.47 (enhanced) \\
MS/Scalar   &  271 (enhanced)  & $2.28$  (positive) & 3.61 (same) \\
MH/Scalar   &  274             &  2.33              & 3.54           \\
MS/Vector   &  268             &  2.59              & 3.33 \\
MH/Vector   &  270             &  3.08              & 3.23 
\\\hline\hline
\end{tabular}
\label{table:v012}
\end{table}

\begin{figure*}[t]
\includegraphics[width=18.0cm]{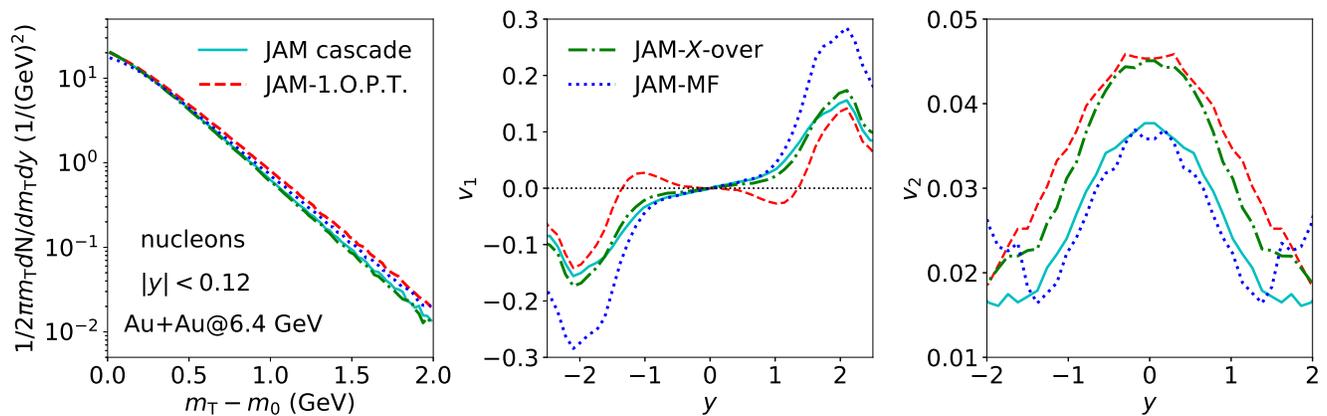}
\caption{Transverse mass distribution (left panel),
$v_1$ (middle-panel),
and $v_2$ (right panel)
for nucleons
in mid-central Au+Au collision at $\srtNN=6.4$ GeV.
Solid, dashed, dashed-doted, and dotted lines show the results from
JAM cascade, JAM with first-order phase transition,
JAM with crossover EoS,
and JAM with hadronic mean field simulations,
respectively.
Spectator nucleons are excluded in the calculations.
}
\label{fig:mtv1v2}
\end{figure*}

The effects of the EoS on the three observables are summarized
in Table~\ref{table:v012}:
The slope parameter, $T_\textrm{eff}$, is extracted from the fit
of the transverse mass $m_\text{T}$
distribution of nucleons at mid-rapidity, $|y|<0.12$,
and $m_\text{T}-m_0 > 0.5$ GeV,
by the exponential function
\begin{equation}
 \frac{1}{2\pi m_\textrm{T}}\frac{dN}{dm_\textrm{T}dy}
 \sim \exp\left(-\frac{m_\textrm{T}}{T_\textrm{eff}}\right).
\end{equation}
The slope of the directed flow measure $F=dv_1/dy$ is extracted
by fitting the rapidity dependence
of $v_1$ by the cubic equation $v_1(y)=Fy+Cy^3$ in the rapidity interval
$|y|<0.8$.
One can see the distinct features of the effects of the EoS on the flows.
Thus, these calculations demonstrate that the combined analysis of
radial, directed and elliptic flows provide the wanted information
on the effect of the EoS/the early pressure of the system.

The dependence of the flow on the EoS as computed
in the hadronic mean field mode as well as for the different
implementations of the potentials are compared
in Table~\ref{table:v012} 
for the different incompressibility values, $K=270$ and 370 MeV,
as well as for the different assumptions of the potentials;
scalar or vector type:
Radial flow is barely influenced
by the details of the model input,
the hadronic EoS dependence is rather small at this beam energy.
Vector potentials predict
larger $v_1$ values and smaller $v_2$ values as compared
to the values obtained in the simulations by the scalar potentials.
However, note the weak dependence of
the different modelings and parameters - this leaves our conclusions
unchanged.

The rapidity dependence of $v_1$ and $v_2$
is investigated in Fig.~\ref{fig:mtv1v2}, which shows
the transverse mass distribution at mid-rapidity, $|y|<0.12$,
the rapidity dependence of $v_1$ and $v_2$
of the nucleons,
for mid-central Au+Au collision at $\srtNN=6.4$ GeV.
Note the characteristic predictions for
all scenarios, not only in the mid-rapidity region, but also
the forward/backward rapidities.
The enhancement of $v_2$ at large rapidities in the mean field simulation
is due to the enhancement of the directed flow $v_1$
at forward/backward rapidities~\cite{Stoecker:1981pg,Sorge:1996pc}.

\section{Conclusions}
\label{sec:conclusion}

The EoS dependence of the excitation function
of the elliptic flow of protons and charged particles
is investigated 
within the microscopic JAM transport approach.
The conjecture originally made by Sorge~\cite{Sorge:1996pc}
has successfully pushed in new directions 
by explicit simulations which
employ different consistent scenarios
for both the EoS and the hadronic scattering terms
in the high baryon density beam energy region
at $\srtNN=3-30$ GeV.
The measurement of radial and elliptic flows
provide information on both, the early and the late pressure.
The proof that $v_2$ is highly sensitive to the EoS is done by
explicitly taking into account the effects of both, a first order
phase transition (or crossover transition), and the hadronic mean fields.
Adding measurements of directed flow shows that
the combined analysis of all three flow coefficients,
the radial, the directed and the elliptic flows allows now
to extract the EoS in different expansion scenarios of the system. 
Hence, experiments to constrain the EoS are doable
and the search for a first order phase transition at high baryon densities
will be taken up in the near future.
 
If the compressed matter does undergo a first order phase transition,
then the considerable enhancement of both,
radial and elliptic flow, are predicted.
Simultaneously, a sudden occurrence of the anti-flow of proton's
$v_1$, to values of $v_1$ less than zero,
proves to be a unique feature
of a first-order phase transition at the highest baryon matter densities.
Hadronic degrees of freedom enhance the radial flow,
but do not enhance the elliptic flow.
Without a first order phase transition,
only a positive slope of the directed flow $v_1$ can be achieved.
Moreover, it is possible to distinguish between a first order phase
transition and a crossover:
A crossover does not enhance the radial flow,
enhances the elliptic flow, but yields a \textit{positive} slope of
the directed flow.
These clear differences do allow experimentally to prove whether
a first-order phase transition did occur, using simultaneously 
the radial, the directed and the elliptic flow values.
These independent sensitivities to the EoS
constitute a unique way to extract firmly the properties of
high density baryon matter.
In the future,
a study of
HBT correlations as well as the study of fluctuations of conserved quantities 
can probe the different expansion dynamics,
and search for the conjectured critical point of the EoS.
Systematic studies of the centrality dependence, and
of various identified particle species's spectrum 
provide information on heat conduction coefficients,
bulk- and shear viscosity values,
on top of the EoS dependence of the high density baryon matter
created in high energy heavy-ion collisions.
The EoS at finite baryon density is still unknown.
Therefore,
different EoS should be tested experimentally.
For example, it will be interesting to use
an EoS based on lattice QCD data, e.g. via the PDPL$\chi$RMF~\cite{CMEOS}
and the Quantum van der Waals (QvdW) model~\cite{Vovchenko:2016rkn}.
A non-trivial structure
is dynamically formed in the chiraly symmetric phase
by a non-equilibrium time evolution of the chiral $\sigma$-field,
which is not correlated to the reaction plane,
leading to a reduction of elliptic flow
in the case of  a first order phase transition~\cite{Paech:2003fe}.
As a next step, a non-equilibrium real-time simulation,
which treats a phase transition
dynamically will be considered.

Future experiments currently planned like,
BES II of STAR at RHIC~\cite{BESII},
the Compressed Baryonic Matter (CBM) experiment at FAIR~\cite{FAIR},
MPD at NICA,  JINR~\cite{NICA},
and a heavy ion experiment at J-PARC (J-PARC-HI)~\cite{Sako:2014fha}
offer opportunities 
at the most favourable beam energies
to explore the highest baryon density matter,
and to study the phase structure of  QCD.

\section*{Acknowledgement} 
H.~S.~ thanks Nu Xu for numerous useful discussions.
Y. N. thanks the Frankfurt Institute of Advanced Studies where part of this
work was done.
This work was supported in part by the
Grants-in-Aid for Scientific Research from JSPS
(Nos.15K05079,  
15K05098, 
and 17K05448).
H.~N.~ has received funding from the European Union's Horizon 2020 research and
innovation programme under the Marie Sklodowska-Curie grant agreement no.
655285 and from the Helmholtz International Center for FAIR within the
framework of the LOEWE program launched by the State of Hesse.
H. St.~ appreciates the generous endowment of the 
 Judah M. Eisenberg Laureatus professorship.
X. Luo is supported in part by the MoST of China 973-Project No.2015CB856901
and NSFC under grant No. 11575069.
Computational resources have been provided by the Center for Scientific
Computing (CSC) at the J. W. Goethe-University, Frankfurt,
and GSI, Darmstadt.

\end{document}